%% file: make_astro.tex
\documentclass{mpe_report}

\usepackage{psfig,graphicx,epsfig}
\usepackage{color}
\usepackage{amsmath,amssymb,epic,eepic,array}

\begin{document}

\pagenumbering{arabic}
\setcounter{page}{92}

\renewcommand{\FirstPageOfPaper }{ 92}\renewcommand{\LastPageOfPaper }{ 95}\include{./mpe_report_weltevredep}  \clearpage

\end{document}

%% file: mpe_report_weltevredep.tex

\newcommand{\degrees}[1]{\ensuremath{#1^\circ}}

\def\NrPulsars{187 } 
\def\NrNewDrifters{42}
\def\NrDrifters{68 }
\def\NrPulsarsSNR{106 }
\def\NrDriftersSNR{57 }
\def\DriftPercentageSNR{54}


\title{Drifting subpulse survey at 21 cm}
\author{P. Weltevrede\inst{1} \and  R.~T. Edwards\inst{1,3} \and B.~W. Stappers\inst{2,1}}
\institute{Astronomical Institute ``Anton Pannekoek'', University of Amsterdam, 
Kruislaan 403, 1098 SJ Amsterdam, The Netherlands \and
Stichting ASTRON, Postbus 2, 7990 AA Dwingeloo, The Netherlands \and
CSIRO Australia Telescope National Facility, PO Box 76, Epping NSW 1710,  Australia
}
\maketitle

\begin{abstract}
We present the statistical results of a systematic, unbiased search
for subpulse modulation of \NrPulsars pulsars performed with the
Westerbork Synthesis Radio Telescope (WSRT) in the Netherlands at an
observing wavelength of 21 cm (Weltevrede et~al.~ 2006a). We have
increased the list of pulsars that show the drifting subpulse
phenomenon by \NrNewDrifters, indicating that more than 55\% of the
pulsars that show this phenomenon. The large number of new drifters we
have found allows us, for the first time, to do meaningful statistics
on the drifting phenomenon. We find that the drifting phenomenon is
correlated with the pulsar age such that drifting is more likely to
occur in older pulsars. Pulsars that drift more coherently seem to be
older and have a lower modulation index. Contrary claims from older
studies, both $P_3$ (the repetition period of the drifting subpulse
pattern) and the drift direction are found to be uncorrelated with
other pulsar parameters.
\end{abstract}

\begin{figure}[tb]
\begin{center}
\centerline{\psfig{file=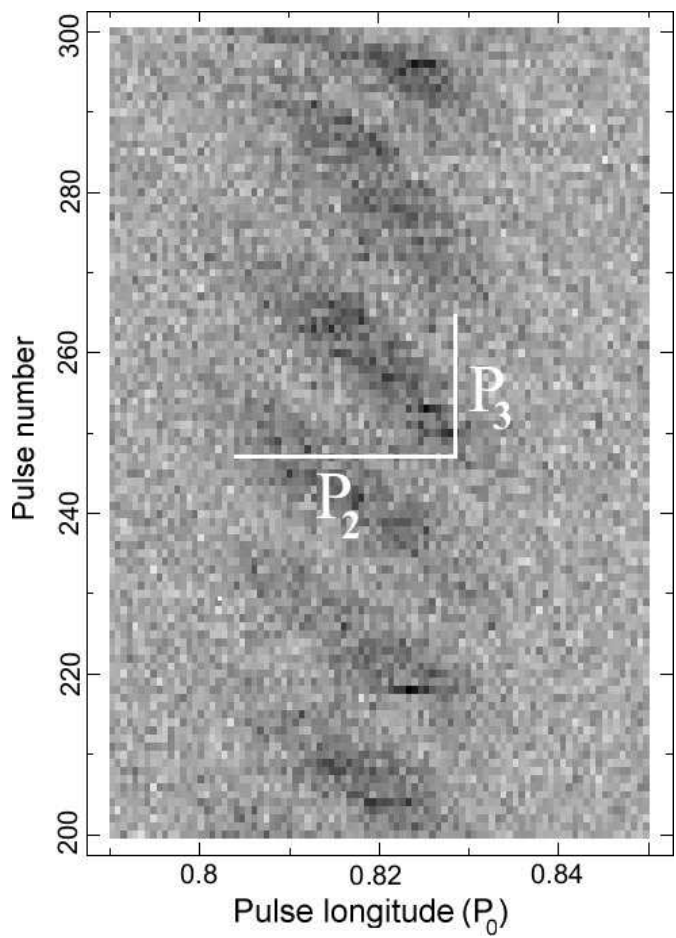,width=8.8cm,clip=}}
\end{center}
\caption{\label{stack} A pulse-stack of one hundred successive pulses
of PSR B1819$-$22, which was discovered to have drifting
subpulses. The drift bands can be characterized by the values $P_2$
and $P_3$.}
\end{figure}

\section{Introduction}

If one can detect single pulses one can see that in some pulsars they
consist of subpulses and in some cases these subpulses drift in
successive pulses in an organized fashion through the pulse window. If
one plots a so-called ``pulse-stack'', a plot in which successive
pulses are displayed on top of one another, the drifting phenomenon
causes the subpulses to form ``drift bands''. In Fig. \ref{stack} one
can see a sequence of 100 pulses of a pulsar which was discovered to
have clear drifting subpulses.  The pulse number is plotted vertically
and the time within the pulses (i.e. the pulse longitude)
horizontally. The drift bands are characterized by two numbers: the
horizontal separation between them in pulse longitude ($P_2$) and the
vertical separation in pulse periods ($P_3$).  This complex, but
highly regular intensity modulation in time is known in great detail
for only a small number of well studied pulsars. Because the
properties of the subpulses are most likely determined by the emission
mechanism, we learn about the physics of the emission mechanism by
studying them. That drifting is linked to the emission mechanism is
suggested by the fact that drifting is affected by ``nulls'', where
nulling is the phenomenon whereby the emission mechanism switches off
for a number of successive pulses. The main goals of this unbiased
search for pulsar subpulse modulation is to determine what percentage
of the pulsars show the drifting phenomenon and to find out if these
drifters share some physical properties. As a bonus of this
observational program new, individually interesting drifting, subpulse
systems are found (Weltevrede et~al.~ 2006a, 2006b, 2006c).


\section{Observations and data analysis}

An important aspect when calculating the statistics of drifting is
that one has to be as unbiased as possible, so we have selected our
sample of pulsars based only on the predicted signal-to-noise ($S/N$)
ratio in a reasonable observing time. While this sample is obviously
still luminosity biased, it is not biased towards well-studied
pulsars, pulse profile morphology or any particular pulsar
characteristics as were previous studies.
Moreover, all the
conclusions in this paper are based on observations at a single
frequency.  All the analyzed observations were collected with the WSRT
in the Netherlands at an observation wavelength of 21 cm.

One basic method to find out if there is subpulse modulation is to
calculate the modulation index, which is a measure of the factor by
which the intensity varies from pulse to pulse and could therefore be
an indication for the presence of subpulses. To determine if the
subpulses are drifting, the Two-Dimensional Fluctuation Spectrum
(2DFS; Edwards \& Stappers~\cite{es02}) is calculated. By analyzing
the 2DFS it can be determined if this modulation is disordered or
(quasi-)periodic and if there exists a systematic drift. The
calculation of the 2DFS is an averaging process and this makes it a
powerful tool to detect drifting subulses, even when the $S/N$ is too
low to detect single pulses.  The drifting is classified as {\em
coherent} when the drift has a well defined $P_3$ value. For more
details about the observations and data analysis we refer to
Weltevrede et~al. 2006a).

\begin{figure}[tb]
\centerline{\psfig{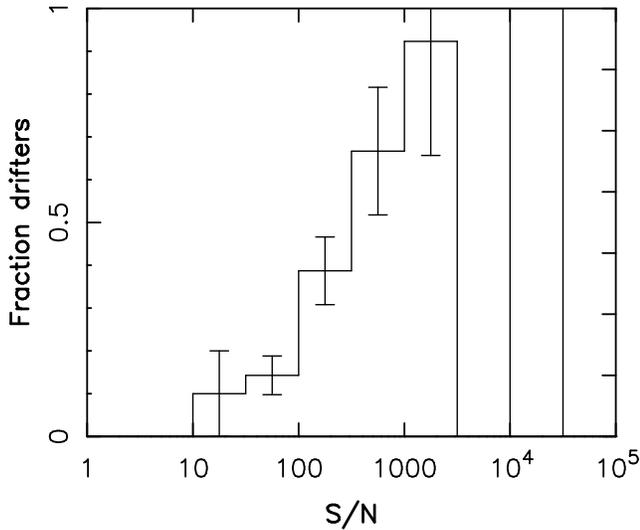}}
\caption{\label{s2n} The fraction
of pulsars we observe to show the drifting phenomenon versus the
measured $S/N$ ratio of the observation.}
\end{figure}

\section{Statistics}

\subsection{The numbers}

Our sample of pulsar is not biased on pulsar type or any particular
pulsar characteristics. This allows us, first of all, to address the
very basic question: what fraction of the pulsars show the drifting
phenomenon?  Of the \NrPulsars analyzed pulsars \NrDrifters pulsars
show the drifting phenomenon, indicating that at least one in three
pulsars drift. This is a lower limit for a number of reasons. First of
all, not all the observations have the expected $S/N$. This could be
because of radio interference, interstellar scintillation,
digitization effects, or because the flux or pulse width for some
pulsars was wrong in the database used.

In the Fig. \ref{s2n} the fraction of pulsars that show the drifting
phenomenon is plotted versus the $S/N$ ratio of the observation. One
can see that the probability of detecting drifting is higher for
observations with a higher $S/N$.  To make the statistics more
independent of the $S/N$ ratio of the observations, the statistics are
done with the \NrPulsarsSNR pulsars with a $S/N \geq 100$. Of these
pulsars \DriftPercentageSNR\% is detected to be drifters and from
Fig. \ref{s2n} it is clear that the real drift percentage could even
be higher.  There are many reasons why drifting is not expected to be
detected for all pulsars. For instance for some pulsars the line of
sight cuts the magnetic pole centrally and therefore longitude
stationary subpulse modulation is expected. Also, refractive
distortion in the pulsar magnetosphere or nulling will disrupt the
drift bands, making it difficult or even impossible to detect
drifting. Some pulsars are known to show organized drifting subpulses
in bursts. In that case (or when $P_3$ is very large) some of our
observations could be too short to detect the drifting.

\begin{figure*}[tb]
\centerline{\psfig{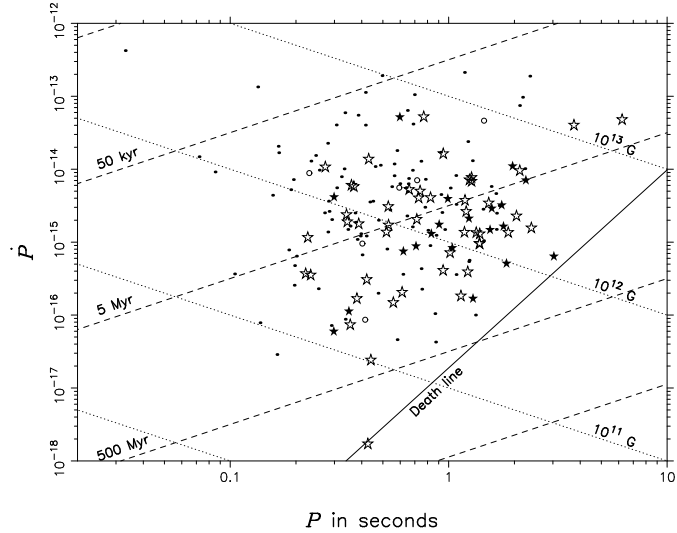}}
\caption{\label{ppdot}The $P$-$\dot P$ diagram of the analyzed pulsars
(including the low $S/N$ observations), where $P$ is the pulse period
and $\dot P$ its time derivative. The non drifting pulsars are the
dots, the coherent drifters are the filled stars, the pulsars with
less regular drifting subpulses are the open stars, and the pulsars
showing longitude stationary subpulse modulation are the open
circles. Lines of equal surface magnetic field strength and
characteristic ages are plotted, as well as a death line. The
millisecond pulsars are not plotted to make the plot more readable.}
\end{figure*}

With a lower limit of one in two it is clear that drifting is at the
very least a common phenomenon for radio pulsars.  This is consistent
with the conclusion that the drifting phenomenon is only weakly
correlated with (or even independent of) magnetic field strength
(see Fig. \ref{ppdot}), because the drifting phenomenon is
too common to require very special physical conditions. It could well
be that the drifting phenomenon is an intrinsic property of the
emission mechanism although for some pulsars it is difficult or even
impossible to detect.

\subsection{The age dependence of the drifting phenomenon}

Two directly measurable and therefore important physical parameters of
the pulsar are the pulse period and its time derivative (spin-down
parameter). From the $P$-$\dot P$ diagram (Fig. \ref{ppdot}) it can be
seen that the population of pulsars that show the drifting phenomenon
(the stars) is on average older than the population of pulsars that do
not show drifting (the dots).
Moreover it seems that the population of pulsars that show coherent
drifting (the filled stars) are on average older than those who show
less regular drifting subpulses (the open stars).
It turns out that the drifters and nondrifters have significantly
different age distributions and that the pulsars which drift
coherently are likely to have a separate age distribution.
It is intriguing to think that drifting
becomes more and more coherent for pulsars with a higher age. A
possible mechanism to distort the drift bands is nulling. However it
has been found that the nulling fraction is on average higher for
older pulsars, showing that nulling cannot explain this correlation.

Another possible scenario is that the alignment of the magnetic dipole
axis with the rotation axis has something to do with the observed
trend. Observations seem to show that the angle $\alpha$ between the
magnetic axis and the rotation axis is on average smaller for older
pulsars and this angle is likely to be an important physical parameter
in the mechanism that drives the drifting phenomenon.  In this
scenario as the pulsar gets older, the rotation axis and the magnetic
axis grows more aligned, which makes the drifting mechanism more
effective or regular.  Also the pulse profile morphology seems to
evolve when the pulsar ages what could make drifting subpulses more
likely to be detected in older pulsars. In the non-radial pulsations
model this trend can also be explained, because the appearance of
narrow drifting subpulses is favored in pulsars with an aligned
magnetic axis (Clemens \& Rosen~\cite{cr04}).

\subsection{The Drifting Phenomenon And The Modulation Index}
\label{ModulationSection}

The drifting phenomenon is a form of subpulse modulation, so the
modulation index is an obvious parameter to try to correlate with the
drifting phenomenon.  There is possably a trend that pulsars that show
the drifting phenomenon more coherently have on average a lower
modulation index (not shown to be statistically significant).

To explain the trend, pulsars that drift coherently must either have
on average more subpulses per pulse or the subpulse intensity
distribution must be more narrow.  The latter could be understood
because coherent drifting could indicate that the electrodynamical
conditions in the sparking gap are stable. Also the presence of
subpulse phase steps results in a lower modulation index and could be
explained as the result of interference between two superposed
drifting subpulse signals that are out of phase (e.g. Edwards \&
Stappers~\cite{es03c}). It is not unlikely that this interference can
only occur if the drifting is coherent, which could explained the
trend. It is also found that many pulsars must have a non-varying
component in their emission, consistent with the presence of
superposed out of phase subpulse signals.  Another explanation for
this trend would be that for some pulsars the organized drifting
subpulses are more refractively distorted than for others, causing the
subpulses to appear more disordered in the pulse window. Moreover it
could be expected that the intensities of the individual subpulses
varies more because of lensing (e.g. Petrova~\cite{pet00}) and
possible focusing of the radio emission (Weltevrede
et~al.~\cite{wsv+03}), causing the modulation index to be higher in
those pulsars.

The modulation index of core type emission is observed to be in
general lower than that of conal type of emission. This is also a
consequence of the Gil \& Sendyk (\cite{gs00}) model. In the sparking
gap model, the drifting phenomenon is associated with conal emission
and therefore expected to be seen in pulsars with an on average higher
modulation index. If well organized coherent drifting is an
exclusively conal phenomenon, it is expected that coherent drifters
have an on average a higher modulation index, exactly opposite to the
observed trend. No drifting is expected for pulsars classified as
"core single stars". Although this may be true for many cases there
are some exceptions, stressing the importance of being unbiased on
pulsar type when studying the drifting phenomenon.

In the framework of the sparking gap model the subpulses are generated
(indirectly) by discharges in the polar gap (i.e. sparks). The number
of sparks that fits on the polar cap is quantified by the complexity
parameter (Gil \& Sendyk~\cite{gs00}), which is expected to be
anti-correlated with the modulation index (Jenet \&
Gil~\cite{jg03}). The complexity parameter is a function of the pulse
period and its derivative and its precise form depends on the model
one assumes for the pulsar emission. By correlating the modulation
index of a sample of pulsars with various complexity parameters as
predicted by different emission models one could try to distinguish
which model best fits the data. We have correlated the modulation
indices in our sample of pulsars with the complexity parameter of four
different emission models as derived by Jenet \& Gil
(\cite{jg03}). Unfortunately none of the models can be ruled out based
on these observations.

\subsection{Properties of the drift behavior}

A significant correlation between $P_3$ and the pulsar age has been
reported in the past (e.g. Rankin~\cite{ran86}). 
However, there is no such correlation found in our data There is also
no correlation found between $P_3$ and the magnetic field strength or
the pulse period as well as between the drift direction and the pulsar
spin-down as reported in the past (Ritchings \& Lyne~\cite{rl75}).
The evidence for a pulsar subpopulation located close to the
$P_3=2P_0$ Nyquist limit also seems to be weak.

In a sparking gap model one would expect that the spark-associated
plasma columns drift because of an $\mathbf{E}\times \mathbf{B}$
drift, which depends on both the pulse period and its derivative.  The
absence of any correlation between $P_3$ and a physical pulsar
parameter is difficult to explain in this model, unless many pulsars
in our sample are aliased. Because the emission entities are only
sampled once per rotation period of the star, it is very difficult to
determine if the subpulses in one drift band correspond to the same
emission entity for successive pulses. For instance for PSR B1819$-$22
(see Fig. \ref{stack}) we do not know if the emission entities drift
slowly toward the leading part of the pulse profile (not aliased) or
faster toward the trailing part of the pulse profile (aliased).  If a
pulsar is aliased, a higher $\mathbf{E}\times \mathbf{B}$ drift can
result in a lower $P_3$ value and visa versa, making $P_3$ not a
direct measure of the $\mathbf{E}\times \mathbf{B}$ drift.  Also if
$P_2$ is highly variable from pulsars to pulsar, any correlation with
$P_3$ is expected to be weaker.

\section{Conclusions}
The number of pulsars that are known to show the drifting phenomenon
is significantly expanded by 42 and the fraction of pulsars that show
the drifting phenomenon is likely to be larger than 55\%. This implies
that the physical conditions required for the drifting mechanism to
work cannot be very different than the required physical conditions
for the emission mechanism of radio pulsars. It could well be that the
drifting phenomenon is an intrinsic property of the emission
mechanism, although drifting could in some cases be very difficult to
detect.

Our results seem to suggest that drifting is not exclusively related
to conal emission. 
Although significant correlations between $P_3$ and the pulsar age,
the magnetic field strength and the pulse period have been reported,
we find no such correlations in our enlarged sample. The absence of a
correlation between $P_3$ and any physical pulsar parameter is
difficult to explain, unless many pulsars in our sample are aliased or
if $P_2$ is highly variable from pulsar to pulsar.

The population of pulsars that show the drifting phenomenon are on
average older than the population of pulsars that do not show drifting
and it seems that drifting is more coherent for older pulsars. The
evolutionary trend found seems to suggest that the mechanism that
generates the drifting subpulses gets more and more stable as the
pulsar ages.

If subpulse phase steps are exclusively (or at least more likely) to
occur in pulsars with coherently drifting subpulses, their modulation
index is expected to be on average lower. This is indeed the trend the
we observe. Another possible scenario to explain this trend is that
coherent drifting indicates that the electrodynamical conditions in
the sparking gap are stable or that refraction in the magnetosphere is
stronger for pulsars that do not show the drifting phenomenon
coherently.


%% file: make_astro.bbl
\begin{thebibliography}{} 
\bibitem[2004]{cr04}{Clemens}, J.~C. \& {Rosen}, R. 2004, ApJ, 609, 340
\bibitem[2003]{jg03}{Jenet}, F.~A. \& {Gil}, J. 2003, ApJ, 596, L215
\bibitem[2002]{es02}{Edwards}, R.~T. \& {Stappers}, B.~W. 2002, A\&A, 393, 733
\bibitem[2003]{es03c}{Edwards}, R.~T. \& {Stappers}, B.~W. 2003, A\&A, 410, 961
\bibitem[2000]{gs00}{Gil}, J.~A. \& {Sendyk}, M. 2000, ApJ, 541, 351
\bibitem[2000]{pet00}{Petrova}, S.~A. 2000, A\&A, 360, 592
\bibitem[1986]{ran86}Rankin, J.~M. 1986, ApJ, 301, 901
\bibitem[1975]{rl75}Ritchings, R.~T. \& Lyne, A.~G. 1975, Nature, 257, 293
\bibitem{wes06} {Weltevrede}, P., {Edwards}, R.~T., \& {Stappers}, B.~W., 2006a,  A\&A, 445, 243
\bibitem{wsr+06} {Weltevrede}, P., {Stappers}, B.~W., {Rankin}, J.~M., \& {Wright}, G.~A.~E.  2006b, Astrophys. J., 645, L149
\bibitem[2003]{wsv+03}{Weltevrede}, P., {Stappers}, B.~W., {van den Horn}, L.~J., \& {Edwards}, R.~T.  2003, A\&A, 412, 473
\bibitem{wws+06} {Weltevrede}, P., {Wright}, G.~A.~E., {Stappers}, B.~W., {Rankin}, J.~M., \&  2006c, accepted by A\&A ({\tt astro-ph/0608023}).
\end{thebibliography}
